# Enhanced carrier mobility in anisotropic 2D tetrahex-carbon through strain engineering

Xihong Peng, [1*] Qun Wei, [2] Guang Yang, [1, 3]

[1]College of Integrative Sciences and Arts, Arizona State University, Mesa, Arizona 85212, USA

[2]School of Physics and Optoelectronic Engineering, Xidian University, Xi'an, Shaanxi 710071, P. R. China

[3]School of Science, Hebei University of Science and Technology, Shijiazhuang, Hebei 050018, P. R. China

## ABSTRACT

A recently predicted two dimensional (2D) carbon allotrope, tetrahex-carbon consisting of tetragonal and hexagonal rings, draws research interests due to its unique mechanical and electronic properties. Tetrahex-C shows ultrahigh strength, negative Poisson's ratio, a direct band gap and high carrier mobility. In this work, we employ first-principles density-functional theory calculations to explore the directional dependence of electronic properties such as carrier effective mass and mobility in tetrahex-C. Tetrahex-C demonstrates strong anisotropicity in effective mass of charge carrier and therefore its mobility (electric conductance) exhibits a strong orientation preference. More interesting, we find that such unique anisotropic carrier effective mass and mobility can be controlled by simple uniaxial strain. The orientation dependence of effective mass can be dramatically rotated by 90° through applying uniaxial tensile strain beyond ~ 7% (11%) in the armchair direction for the hole (electron). As a result, the intrinsic carrier mobility in tetrahex-C is significantly enhanced. The results are useful for potential electronic and mechanical applications in tetrahex-C.

1. **Introduction**

Two dimensional (2D) structures such as graphene [1–3], transition metal dichalcogenides (TMDs) [4–7], and phosphorene [8,9] were successfully fabricated in lab and prompted remarkable research interests in 2D materials. Graphene was considered to have potential applications in future electronics [1–3] due to its unique properties such as ultrahigh carrier mobility $10^4 \sim 10^6$ cm$^2$/(V·s) [1,10–12] depending on situations of substrates and impurity level. However, the zero-band-gap of pristine graphene limits its electronic applications. Other 2D materials such as TMDs [4–7] and phosphorene [8,9] possess the virtue of a finite direct band gap, a significant advantage over graphene for applications in optoelectronic and electronic devices. MoS$_2$ retains carrier mobility $\sim 200$ cm$^2$/(V·s) [13]. The hole mobility of phosphorene is $\sim 10^4$ cm$^2$/(V·s) [14].

A new 2D carbon allotrope, tetrahex-carbon consisting of squares and hexagons, was predicted by Ram and Mizuseki [15] based on the structure of penta-graphene [16]. This new structure possesses slightly lower energy than penta-graphene and implies larger opportunity to be fabricated in lab. It shows a direct band gap at Γ (HSE gap $\sim 2.6$ eV) with high electron mobility $\sim 10^4$ cm$^2$/(V·s) [15]. Through first-principles calculations, it was found that this material exhibits ultrahigh ideal strength outperforming both graphene and penta-graphene and demonstrates intrinsic negative Poisson's ratio [17]. In addition, tetrahex-C remains integrity of direct-gap with strain. It was found that the direct-band-gap feature remains intact up to 16.4% of biaxial strain [15] and 20% uniaxial strain in the zigzag direction [17], unlike TMDs and phosphorene which experience direct-to-indirect band gap transition upon strain applications [9,18–21].

Tetrahex-C also possesses remarkable anisotropic behavior, especially in carrier effective mass and mobility. In this work, we find the effective mass of electron has a value of 0.23 $m_e$ versus 1.77 $m_e$ in the zigzag and armchair directions, respectively. The directional dependence of hole effective mass is even more astonishing with 13.88 $m_e$ in the zigzag while 0.34 $m_e$ in the armchair direction. The anisotropicity in carrier effective mass indicates a strong orientation preference in the carrier mobility (and electric conductance). More fascinating, we find that this exceptional anisotropic effective mass and mobility can be controlled by simple uniaxial strain. The anisotropicity in effective mass (thus the orientation preference in electric conductance) can be dramatically rotated by 90° through applying uniaxial tensile strain beyond $\sim 7\%$ in the armchair direction. This effect is summarized in a schematics in Fig. 1(c). As a result, the carrier mobility can be largely enhanced in some direction.

2. **Computational methods**

The first-principles density-functional theory (DFT) [22] calculations are performed using the Vienna *ab initio* simulation package (VASP) [23,24]. The Perdew-Burke-Ernzerhof (PBE) exchange-correlation functional [25] is chosen for general electronic structure calculations and geometry relaxation. The hybrid Heyd-Scuseria-Ernzerhof (HSE)06 method [26,27] is used to calculate electronic band structures for a better prediction on the band gap of the system. The projector-augmented wave (PAW) [28,29] potentials are used to treat interactions between ion cores and valance electrons. Plane waves are used to expand the valance electron ($2s^2 2p^2$) wavefunctions. The reciprocal space is meshed using Monkhorst-Pack method. The plane wave kinetic energy cutoff 900 eV and 15 × 13 × 1 mesh for reciprocal space are chosen in geometry relaxation and force fields calculations along with the PBE functional. The energy convergence criterion for electronic iterations is set to be $10^{-6}$ eV and the force is converged within 0.001 eV/Å with the energy criterion $10^{-5}$ eV for geometry optimization of the unit cell. The kinetic energy cutoff 500 eV for plane wave basis set is used for the HSE band structure calculations. The *z*-vector of the unit cell is set to be 20 Å to ensure sufficient vacuum space included in the calculations to minimize the interaction between the system and its replicas resulted from the periodic boundary condition.

In the band structure calculations, 31 *k*-points are collected along each high symmetry line in the reciprocal space. Carrier effective mass is defined as $m^* = \hbar^2 (\frac{\partial^2 E}{\partial k^2})^{-1}$, where ℏ is the reduced Plank constant, *E* is the energy of conduction band for electron or valance band for hole, *k* is the reciprocal wave vector. The effective mass of charge carrier is evaluated through a five-point second derivative, considering a *k*-point spacing smaller than 0.015 Å$^{-1}$ in order to avoid non-parabolic effects.

The initial structure of tetrahex-C is obtained according to the reference [15]. Unlike flat graphene, this 2D carbon network is buckled and consists with tetragonal and hexagonal rings as shown in Fig. 1. The unit cell has 12 carbon atoms, which are either *sp$^2$* or *sp$^3$* hybridized. The *sp$^3$* hybridized atom is sandwiched between two layers of *sp$^2$* bonded carbon. Starting with the optimized structure, uniaxial tensile strain is applied in the *x* (zigzag) or *y* (armchair) direction.

The tensile strain is defined as $\varepsilon = \frac{a-a_0}{a_0}$, where $a$ and $a_0$ are the lattice constants of the strained and relaxed structure, respectively. With strain applied in the *x*- or *y*-direction, the lattice constant in the transvers direction is fully relaxed through minimization of the total energy to ensure no stress in the transverse direction.

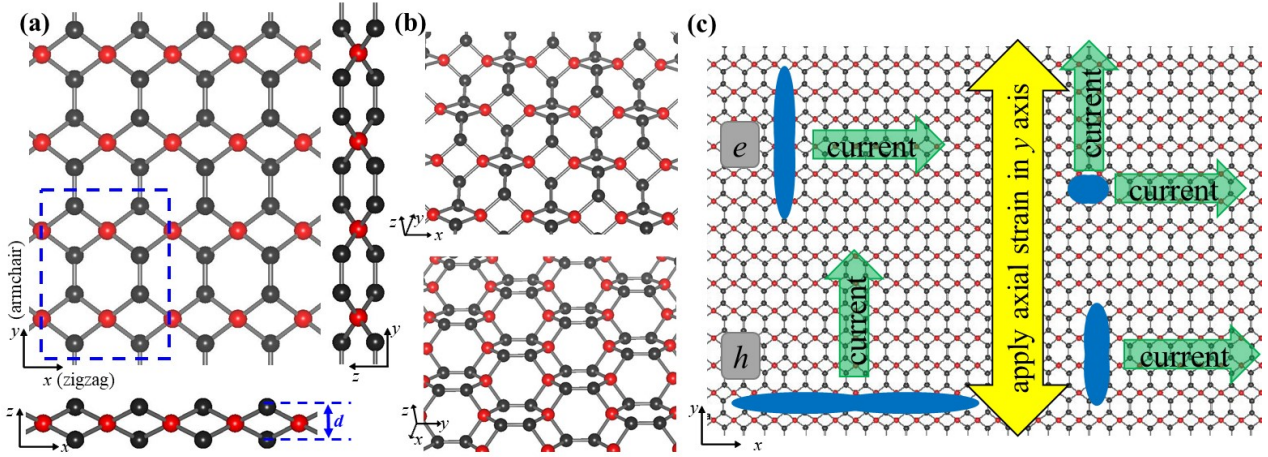

*Figure 1. (a)(b) Snapshots of buckled tetrahex-C. The dashed rectangle in (a) represents a unit cell. The sp² and sp³ hybridized carbon atoms are in black and red, respectively. (c) Schematics of the anisotropicity of effective masses of electron and hole (left) without strain and (right) with axial strain applied in the armchair direction, and the corresponding preferred current direction in tetrahex-C.*

### 3. Results and discussion

#### A. Effective mass of electron and hole in tetrahex-C

Our calculated lattice constants of tetrahex-C are $a$ = 4.531 Å, $b$ = 6.102 Å, buckling thickness $d$ = 1.163 Å, which are in great agreement with literature [15,17]. Tetrahex-C possesses a direct band gap at the center of Brillouin zone [15]. Our calculations confirm this and the band structure is presented in Fig. 3 (a) based on the hybrid HSE method. Our calculated HSE band gap is 2.64 eV, close to the value of 2.63 eV reported in the literature [15].

The carrier effective mass is calculated using $m^* = \hbar^2 (\frac{\partial^2 E}{\partial k^2})^{-1}$, where $\hbar$ is the reduced Plank constant, $E$ is the energy of conduction band for electron or valance band for hole, $k$ is the reciprocal wave vector. We calculate the effective mass in an arbitrary direction along Γ-A with an angle $\theta$ from the $k_x$ direction as shown in Fig. 2(a). The calculated direction-dependence of the effective mass of electron and hole in the relaxed tetrahex-C is presented in Fig. 2(b)-2(c). It is clear that tetrahex-C demonstrates a strong anisotropic feature. The effective mass of electron shows the smallest value of 0.23 $m_e$ in the $x$ (zigzag) direction, while possesses the largest value of 1.77 $m_e$ in the $y$ (armchair) axis. This is also clearly indicated in the electronic band structure as shown in Fig. 3(a). The conduction band (state C) has a steep curve along the $x$ direction while a rather flat band along the $y$ axis, resulting a small (big) effective mass of electron in the $x$ ($y$) direction. However, the situation is opposite for the hole as shown in Fig. 2(c), in which the largest effective mass (13.88 $m_e$) is along the $x$-direction while the smallest value (0.34 $m_e$) is in the $y$-axis. This is also implied by the curvature of the valence band (state B) in Fig. 3(a). Note that our calculated effective mass of electron and hole is not in a good agreement with the reference [15], especially along the $x$-direction. We further use the advanced hybrid HSE method to re-evaluate the values. The HSE method gives similar results as our PBE predicted values.

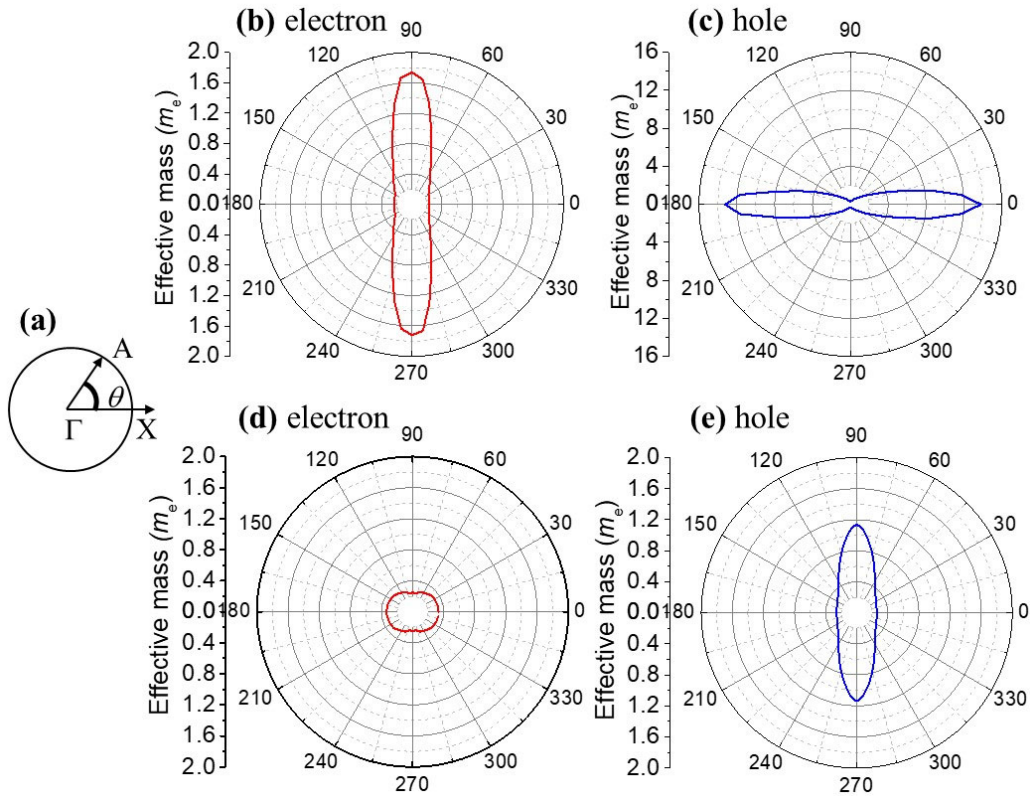

*Figure 2. Directional dependence of carrier effective mass in tetrahex-C. (a) k-vector along an arbitrary direction Γ-A with an angle θ from the $k_x$ direction, effective mass of (b) electron (c) hole as a function of angle θ in the relaxed tetrahex-C. Effective mass of (d) electron (e) hole as a function of angle θ for the structure under 12% strain applied in the armchair direction.*

### B. Uniaxial strain effect on the band structure in tetrahex-C

Uniaxial tensile strain is applied in the zigzag or armchair direction to explore the band structure variation under strain. It is found tetrahex-C maintains phonon stability up to 20% (16%) uniaxial strain along the zigzag (armchair) direction [17]. Under uniaxial tensile strain in the zigzag direction, the energy of the conduction band minimum (CBM, i.e. state C as labeled in Fig. 3(a)) decreases, while that of the valence band maximum (VBM, i.e. state B) increases with strain, resulting a band gap reduction [17]. The direct-gap feature at Γ remains intact with the CBM (VBM) continuing at state C (B) under uniaxial strain in the zigzag direction up to 20% [17]. In the case of axial strain applied in the armchair direction, although the band gap remains direct (or quasi-direct) at Γ till strain up to 16% [17], state energy crossover takes place so that the CBM (VBM) is no longer at state C(B).

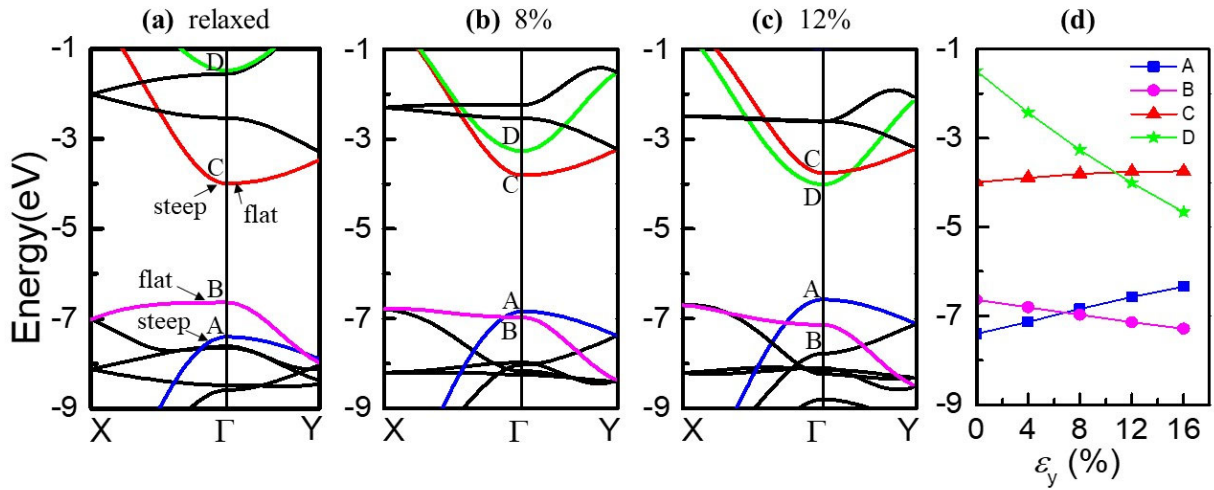

Figure 3. HSE predicted band structure of tetrahex-C. (a) relaxed, (b) 8% and (c) 12% uniaxial strained in the armchair direction, (d) energy of band A-D at Γ as a function of strain. All energies are referenced to vacuum. Energies of states A and B crossover at ~ 7% strain, while states C and D crossover at ~11% strain.

Fig. 3 shows the variation of the band structure with the strain in the armchair direction. For example, at $\varepsilon_y$ = 8% as shown in Fig. 3(b), the VBM is no longer its original state B. Instead, the energy of state A increases rapidly with strain to exceed that of state B, thus represents the new VBM. Therefore, the effective mass of hole should now be evaluated from state A. State A has a steep band along the *x* direction and relatively flat band in the *y* axis, which is opposite to that of state B. Similar case also occurs to the conduction bands. For example, at $\varepsilon_y$ = 12% as shown in Fig. 3(c), the CBM is no longer its original state C. The energy of state D decreases with strain to be lower than that of state C and represents the new CBM. As a result, the effective mass of electron is now evaluated from state D which has a steeper band than state C along the *y* direction to expect a smaller effective mass of electron. The directional dependence of the carrier effective mass in tetrahex-C under strain is presented in Fig. 2(d)-2(e). It shows that the effective mass of electron in the *y*-direction significantly reduced compared to the case in Fig. 2(b). In the case of hole, the direction-dependence of effective mass rotates by 90°, when comparing Fig. 2(c) and 2(e).

Through the analysis of the energy variation of states A-D as a function of strain plotted in Fig 3(d), we can determine at which strain the energy crossover of states A and B (or C and D) occurs. It is found that A and B crossover takes place at ~7% strain while C and D crosses at ~11% strain.

### C. Carrier mobility in tetrahex-C

In addition to the relaxed tetrahex-C, the carrier effective mass is also calculated for the strained system with $\varepsilon_y$ = 4%, 8%, 12%, and 16%. The obtained effective mass of electron and hole as a function of strain is presented in Fig. 4(a)-4(b). The electron has larger effective mass in the *y* direction compared to that in the *x* axis with axial strain in the range of 0 ~ 11%. When strain is beyond 11%, the directional dependence of electron effective mass rotates by 90° with a significantly reduced value in the *y* direction (i.e. 1.39 $m_e$ for 8% strain and 0.26 $m_e$ for 12% strain).

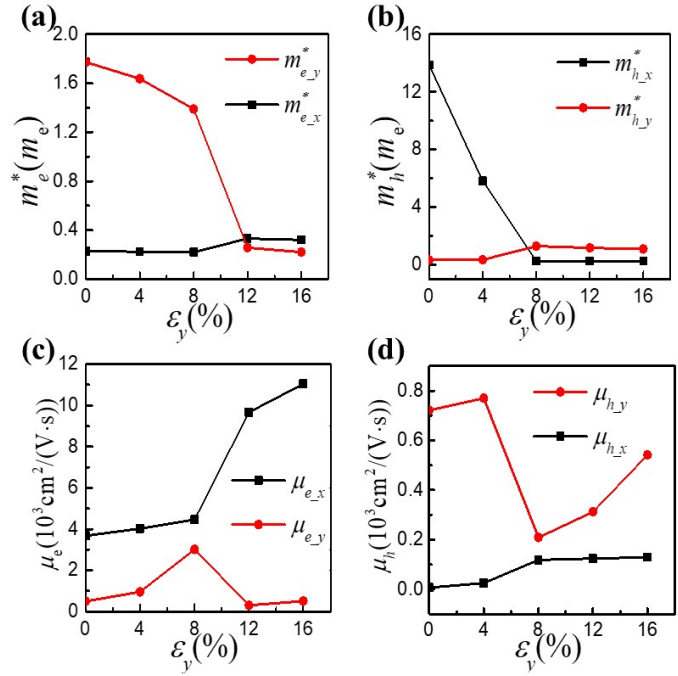

*Figure 4. Effective mass of (a) electron, (b) hole, and carrier mobility of (c) electron, (d) hole in tetrahex-C at room temperature (T = 300 K), as a function of uniaxial strain applied in the armchair direction.*

In the case of hole, the effective mass is larger in the *x* direction than that in the *y* axis in the strain range of 0 ~ 7%. Beyond 7% strain, effective mass of hole dramatically drops from 13.88 $m_e$ in the relaxed structure to 0.25 $m_e$ in the 12% strained one. In the $y$ axis, the hole effective mass increases from 0.34 $m_e$ in the relaxed structure to 1.18 $m_e$ in the 12% strained one. The effective mass values are also listed in Table 1. Our calculated carrier effective mass is checked using the advanced hybrid HSE method. It is found that the HSE predicts the same trends as that with the PBE functional, which further validate the robustness of the PBE results.

The carrier mobility in a 2D system is given by the expression [14,30],

$$\mu_{2D} = \frac{e\hbar^3 C_{2D}}{k_B T m^* m_d (E_1)^2} \quad (1)$$

where $e$ is the elementary charge, $\hbar$ the reduced Plank constant, $k_B$ the Boltzmann's constant, $T$ the temperature ($T$ = 300 K is used in this work), $m^*$ is the effective mass of charge carrier along the transport direction (either $m_x$ or $m_y$ along the $x$ and $y$ direction, respectively), $m_d$ is the equivalent density-of-state mass defined as $m_d = \sqrt{m_x^* m_y^*}$, and $E_1$ is the deformation potential computed through mimicking the lattice deformation due to the carrier-phonon interaction by compressing/dilating the lattice constant and relaxing the primitive cell. It is calculated as $E_1 = \frac{\Delta E}{\Delta l/l_0}$, where $\Delta E$ is the energy change of the band (VBM for hole and CBM for electron) under small lattice compression/dilatation (i.e. $\Delta l/l_0$ = -1%, -0.5%, 0.5%, 1%), $l_0$ is the lattice constant in the transport direction and $\Delta l$ is the deformation of $l_0$, $C_{2D}$ is the elastic constant of the longitudinal strain in the propagation direction ($x$ or $y$) of the longitudinal acoustic wave, which is calculated from the equation $\frac{E-E_0}{A_0} = \frac{C}{2}(\frac{\Delta l}{l_0})^2$, where $E$ and $E_0$ are the energy of the system with and without lattice deformation, $A_0$ is the area of the lattice at equilibrium for the 2D system, $\Delta l$ and $l_0$ are the same meanings as mentioned before.

In order to calculate the elastic constant $C_{2D}$, the strain energy $E - E_0$ as a function of lattice compression/dilation in the $x$ and $y$ directions are plotted in Fig. 5(a) and 5(d), respectively. Through parabolic fitting the curve of the strain energy versus lattice deformation, our obtained elastic constants $C_{2D\_x}$ = 288.6 N/m, $C_{2D\_y}$ = 281.7 N/m, which are consistent with the elastic stiffness constants $C_{11}$ and $C_{22}$ in the references [15,17]. The area $A_0$ of the lattice at equilibrium in tetrahex C is 27.648 Å².

The deformation potential $E_{1\_x}$ and $E_{1\_y}$ for the VBM and CBM are evaluated through lattice dilatation/compression in the $x$ and $y$ direction, respectively. Their energy variations with strain are presented in Fig. 5. The slope of the curve represents the deformation potential. The calculated deformation potential $E_{1\_x}$ is 5.44 (3.36) and $E_{1\_y}$ is 3.38 (3.25) for the VBM (CBM). Our calculated deformation potential values are close to those reported in literature [15].

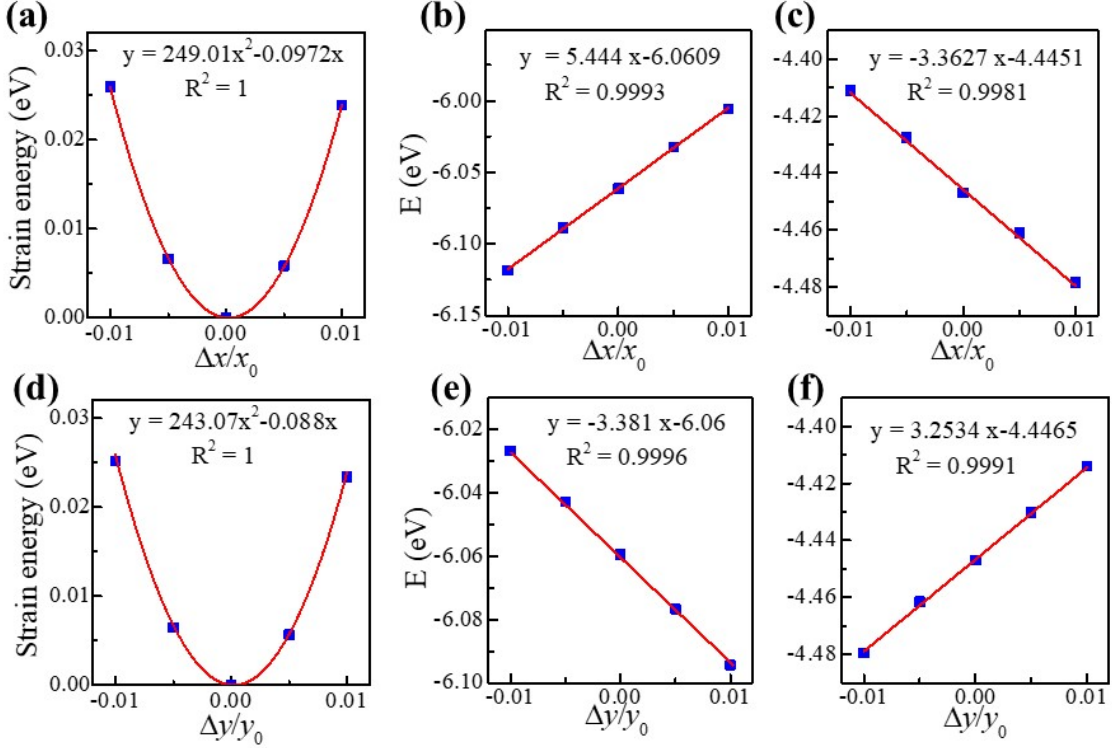

*Figure 5. (a)(d) Strain energy, energy variation of (b)(e) VBM and (c)(f) CBM as a function of lattice dilation/compression near the relaxed structure. Top, lattice dilatation/compression along the x direction; bottom, along the y direction. All state energies of the VBM and CBM are referenced to vacuum.*

The carrier mobility in the relaxed tetrahex-C is $\mu_{e\_x} = 3.682 \times 10^3 \frac{cm^2}{V \cdot s}$, $\mu_{e\_y} = 0.5 \times 10^3 \frac{cm^2}{V \cdot s}$, $\mu_{h\_x} = 0.007 \times 10^3 \frac{cm^2}{V \cdot s}$, $\mu_{h\_y} = 0.721 \times 10^3 \frac{cm^2}{V \cdot s}$. Note that our calculated mobilities are generally smaller than the values reported in Ref. [15]. This is mainly due to the discrepancy of the effective masses of charge carriers along the *x*-direction between our calculation and the reference. Our calculated carrier effective masses in the *x*-direction are bigger, which leads to smaller mobilities.

Since strain demonstrates dramatic effect on tuning effective mass of charger carrier as shown in Fig. 4(a)-4(b), the carrier mobility is expected to be modified by strain. Note that the energy crossover occurs for both CBM and VBM bands with the uniaxial strain applied in the *y*-axis as presented in Fig. 3(d). Therefore, the deformation potentials need to be evaluated near the strained lattice. Fig. 6(a)-6(d) shows the energy variation of the CBM and Fig. 6(f)-6(i) presents that of the VBM as a function of lattice dilation near the strained lattice constant with $\varepsilon_y$ = 4%, 8%, 12%, and 16%, respectively. The slopes in the top and bottom plots give the $E_{1\_y}$ value for the electron and hole, respectively. Fig. 6(a)-6(b) presents the CBM energy for state C and Fig. 6(c)-6(d) for state D. The positive sign of the slope for state C and the negative slope for state D were interpreted according to their corresponding bonding/antibonding characteristics in Ref. [17]. Fig. 6(f) plots the VBM for state B and Fig. 6(g)-6(i) for state A. The negative slope for state B and the positive slope for state A were also elaborated in literature [17].

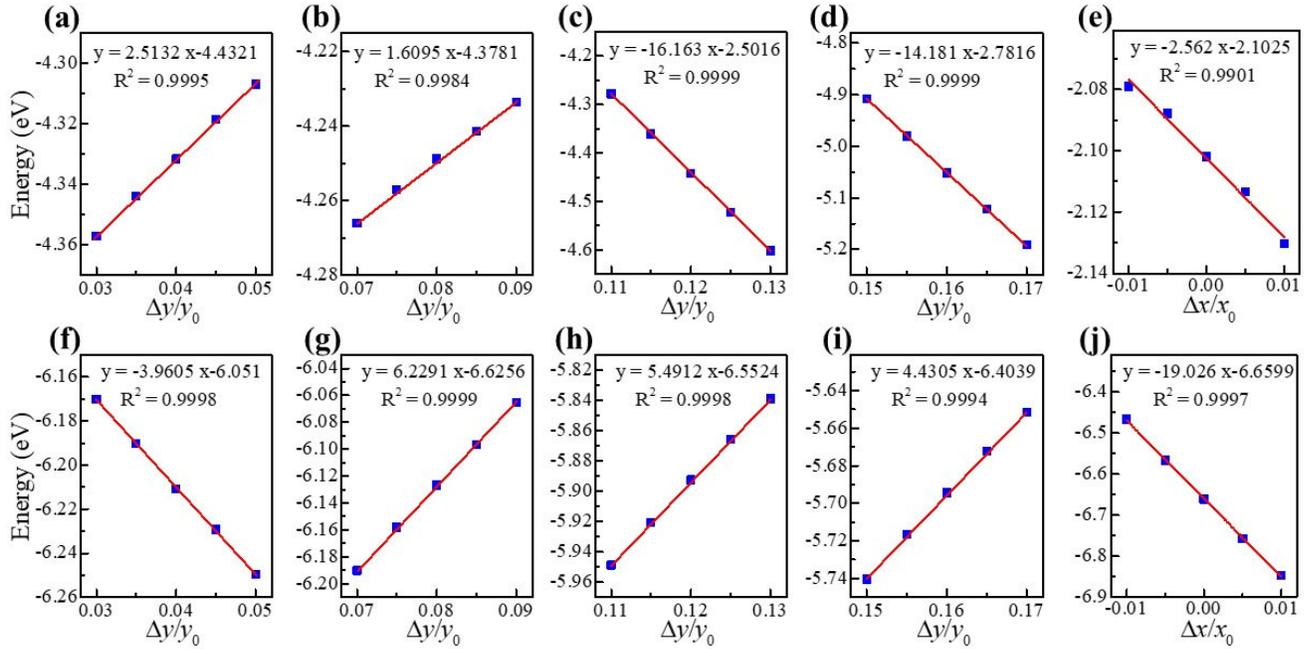

*Figure 6. Energy variation of (a-d) the CBM and (f-i) the VBM as lattice compression/dilation near $\varepsilon_y$ = 4%, 8%, 12% and 16%, respectively. Energy variation of (e) state D and (j) state A as lattice compression/dilation near $\varepsilon_x$ = 0%. The slope of the linear fitting gives the corresponding deformation potential. All state energies are referenced to vacuum.*

To calculate the deformational potential $E_{1\_x}$, the energy variation of the CBM and VBM need to be scanned for the lattice deformation in the x-direction. Since beyond $\varepsilon_y$ ~ 11% strain, state D (instead of the original state C) represents the CBM, the energy variation of state D with lattice deformation needs to be calculated to evaluate $E_{1\_x}$ for the electron and the result is shown in Fig. 6(e). Similarly, beyond $\varepsilon_y$ ~ 7% strain, state A (instead of the original B) represents the VBM, therefore the energy of state A is plotted in Fig. 6(j) to obtain $E_{1\_x}$ for the hole.

The carrier mobility for the strained tetrahex-C can be calculated based on the obtained carrier effective mass and deformation potentials under strain. The results are presented in Fig. 4(c)-4(d) as a function of uniaxial strain in the armchair direction, and also listed in Table 1 along with the charge effective mass and deformation potential for the relaxed and strained tetrahex-C. It is clear that the electron mobility in the x direction $\mu_{e\_x}$ is three times enhanced, from $3.682 \times 10^3 \frac{cm^2}{V \cdot s}$ in the relaxed tetrahex-C to $1.104 \times 10^4 \frac{cm^2}{V \cdot s}$ for the one under strain $\varepsilon_y = 16\%$. The electron mobility in the y direction $\mu_{e\_y}$ shows a peak value $3.017 \times 10^3 \frac{cm^2}{V \cdot s}$ at 8% strain. For the hole, the mobility in the x direction is increased from a minimal value of $7 \frac{cm^2}{V \cdot s}$ in the relaxed structure to $129 \frac{cm^2}{V \cdot s}$ for the 16% strained one. In the y direction, the hole mobility shows a peak value of $961 \frac{cm^2}{V \cdot s}$ at 4% strain. Both electron and hole mobilities can be enhanced with strain engineering in the zigzag or armchair direction.

Table 1. *The calculated carrier effective masses in unit of free-electron mass, deformation potential constants in eV, elastic constants in N/m, carrier mobility in ($10^3$ cm$^2$/(V·s)) along the transport directions.*

| Strain $\varepsilon_y$ | Carrier | $m_x^*$ | $m_y^*$ | $E_{1\_x}$ | $E_{1\_y}$ | $C_{2D\_x}$ | $C_{2D\_y}$ | $\mu_x$ | $\mu_y$ |
|---|---|---|---|---|---|---|---|---|---|
| 0% | electron | 0.23 | 1.77 | 3.36 | 3.25 | 288.6 | 281.7 | 3.682 | 0.500 |
|  | hole | 13.88 | 0.34 | 5.44 | 3.96 | 288.6 | 281.7 | 0.007 | 0.721 |
| 4% | electron | 0.22 | 1.64 | 3.36 | 2.51 | 288.6 | 281.7 | 4.026 | 0.961 |
|  | hole | 5.81 | 0.35 | 5.44 | 3.96 | 288.6 | 281.7 | 0.025 | 0.770 |
| 8% | electron | 0.22 | 1.39 | 3.36 | 1.61 | 288.6 | 281.7 | 4.469 | 3.017 |
|  | hole | 0.25 | 1.30 | 19.03 | 6.23 | 288.6 | 281.7 | 0.118 | 0.209 |
| 12% | electron | 0.33 | 0.26 | 2.56 | 16.16 | 288.6 | 281.7 | 9.669 | 0.306 |
|  | hole | 0.25 | 1.18 | 19.03 | 5.49 | 288.6 | 281.7 | 0.124 | 0.312 |
| 16% | electron | 0.32 | 0.22 | 2.56 | 14.18 | 288.6 | 281.7 | 11.044 | 0.512 |
|  | hole | 0.25 | 1.08 | 19.03 | 4.43 | 288.6 | 281.7 | 0.129 | 0.541 |

4. **Summary**

Through first-principles DFT calculations, we find that the 2D tetrahex-C shows remarkable anisotropic feature in the carrier effective mass, mobility, thus electric conductance. More interesting, this prominent anisotropicity can be controlled by applying a simple uniaxial tensile strain in the armchair direction. The direction-dependence of the effective mass of electron and hole can be dramatically rotated by 90° when the strain is beyond a threshold value 7% (11%) for hole (electron), resulting in an enhanced intrinsic carrier mobility. The electron mobility in the zigzag direction is increased three times from $3.682 \times 10^3 \frac{cm^2}{V \cdot s}$ in the relaxed tetrahex-C to $1.104 \times 10^4 \frac{cm^2}{V \cdot s}$t in the 16% strained one; while the hole mobility in the zigzag direction is increased more than one order of magnitude. The results may suggest potential applications of tetrahex-C in nanomechanics and nanoelectronics.

**Acknowledgement**

This work is financially supported by the Natural Science Foundation of China (Grant No.: 11965005), and the 111 Project (B17035). The authors thank Arizona State University Advanced Computing Center for providing computing resources (Agave and Saguaro Cluster), and the computing facilities at High Performance Computing Center of Xidian University.

* To whom correspondence should be addressed. E-mail: xihong.peng@asu.edu.

**Reference:**

[1]    K. S. Novoselov, A. K. Geim, S. V. Morozov, D. Jiang, Y. Zhang, S. V. Dubonos, I. V. Grigorieva, and A. A. Firsov, Science (80-. ). **306**, 666 (2004).

[2]    K. S. Novoselov, A. K. Geim, S. V. Morozov, D. Jiang, M. I. Katsnelson, I. V. Grigorieva, S. V. Dubonos, and A. A. Firsov, Nature **438**, 197 (2005).

[3]    J. C. Meyer, A. K. Geim, M. I. Katsnelson, K. S. Novoselov, T. J. Booth, and S. Roth, Nature **446**, 60 (2007).

[4]    M. M. Benameur, B. Radisavljevic, J. S. Héron, S. Sahoo, H. Berger, and A. Kis, Nanotechnology **22**, 125706 (2011).

[5]    J. N. Coleman, M. Lotya, A. O'Neill, S. D. Bergin, P. J. King, U. Khan, K. Young, A. Gaucher, S. De, R. J. Smith, I. V. Shvets, S. K. Arora, G. Stanton, H. Y. Kim, K. Lee, G. T. Kim, G. S. Duesberg, T. Hallam, J. J. Boland, J. J. Wang, J. F. Donegan, J. C. Grunlan, G. Moriarty, A. Shmeliov, R. J. Nicholls, J. M. Perkins, E. M. Grieveson, K. Theuwissen, D. W. McComb, P. D. Nellist, and V. Nicolosi, Science (80-. ). **331**, 568 (2011).

[6]    C. Lee, Q. Li, W. Kalb, X. Z. Liu, H. Berger, R. W. Carpick, and J. Hone, Science (80-. ). **328**, 76 (2010).

[7]    C. Feng, J. Ma, H. Li, R. Zeng, Z. Guo, and H. Liu, Mater. Res. Bull. **44**, 1811 (2009).

[8]    F. Xia, H. Wang, and Y. Jia, Nat. Commun. **5**, 4458 (2014).

[9]    H. Liu, A. T. Neal, Z. Zhu, Z. Luo, X. Xu, D. Tománek, and P. D. Ye, ACS Nano **8**, 4033 (2014).

[10] J. H. Chen, C. Jang, S. Xiao, M. Ishigami, and M. S. Fuhrer, Nat. Nanotechnol. **3**, 206 (2008).

[11] F. Chen, J. Xia, D. K. Ferry, and N. Tao, Nano Lett. **9**, 2571 (2009).

[12] S. V. Morozov, K. S. Novoselov, M. I. Katsnelson, F. Schedin, D. C. Elias, J. A. Jaszczak, and A. K. Geim, Phys. Rev. Lett. **100**, 016602 (2008).

[13] B. Radisavljevic, A. Radenovic, J. Brivio, V. Giacometti, and A. Kis, Nat. Nanotechnol. **6**, 147 (2011).

[14] J. Qiao, X. Kong, Z. X. Hu, F. Yang, and W. Ji, Nat. Commun. **5**, 4475 (2014).

[15] B. Ram and H. Mizuseki, Carbon N. Y. **137**, 266 (2018).

[16] S. Zhang, J. Zhou, Q. Wang, X. Chen, Y. Kawazoe, and P. Jena, Proc. Natl. Acad. Sci. U. S. A. **112**, 2372 (2015).

[17] Q. Wei, G. Yang, and X. Peng, ArXiv:1912.10363 (2019).

[18] H. J. Conley, B. Wang, J. I. Ziegler, R. F. Haglund, S. T. Pantelides, and K. I. Bolotin, Nano Lett. **13**, 3626 (2013).

[19] P. Johari and V. B. Shenoy, ACS Nano **6**, 5449 (2012).

[20] Y. Wang, C. Cong, W. Yang, J. Shang, N. Peimyoo, Y. Chen, J. Kang, J. Wang, W. Huang, and T. Yu, Nano Res. **8**, 2562 (2015).

[21] Q. Wei and X. Peng, Appl. Phys. Lett. **104**, 251915 (2014).

[22] W. Kohn and L. J. Sham, Phys. Rev. **140**, A1133 (1965).

[23] G. Kresse and J. Furthmuller, Phys. Rev. B **54**, 11169 (1996).

[24] G. Kresse and J. Furthmuller, Comput. Mater. Sci. **6**, 15 (1996).

[25] J. P. Perdew, K. Burke, and M. Ernzerhof, Phys. Rev. Lett. **77**, 3865 (1996).

[26] J. Heyd, G. E. Scuseria, and M. Ernzerhof, J. Chem. Phys. **118**, 8207 (2003).

[27] J. Heyd, G. E. Scuseria, and M. Ernzerhof, J. Chem. Phys. **124**, 219906 (2006).

[28] P. E. Blöchl, Phys. Rev. B **50**, 17953 (1994).

[29] G. Kresse and D. Joubert, Phys. Rev. B **59**, 1758 (1999).

[30] S. Bruzzone and G. Fiori, Appl. Phys. Lett. **99**, 222108 (2011).